\documentclass[twocolumn,english]{revtex4-1}
\usepackage[T1]{fontenc}
\usepackage[latin9]{inputenc}
\usepackage{units}
\usepackage{amstext}
\usepackage{graphicx}
\usepackage{xcolor}

\makeatletter
\usepackage[caption=false]{subfig}

\@ifundefined{showcaptionsetup}{}{%
 \PassOptionsToPackage{caption=false}{subfig}}
\usepackage{subfig}
\makeatother

\usepackage{babel}
\begin{document}

\title{{\normalsize{}Origin of slow stress relaxation in the cytoskeleton}}
\begin{abstract}
{\normalsize{}Dynamically crosslinked semiflexible biopolymers such
as the actin cytoskeleton govern the mechanical behavior of living
cells. Semiflexible biopolymers nonlinearly stiffen in response to
mechanical loads, whereas the crosslinker dynamics allow for stress
relaxation over time. Here we show, through rheology and theoretical
modeling, that the combined nonlinearity in time and stress leads
to an unexpectedly slow stress relaxation, similar to the dynamics
of disordered systems close to the glass transition. Our work suggests
that transient crosslinking combined with internal stress can explain prior reports of soft glassy rheology of cells, 
in which the shear modulus increases weakly with frequency.}{\normalsize\par}
\end{abstract}

\author{{\normalsize{}Yuval Mulla$^{1}$, F. C. MacKintosh$^{2,3,4*}$, Gijsje
H. Koenderink$^{1}$}}
\email{Corresponding authors: fcmack@gmail.com, g.koenderink@amolf.nl}

\affiliation{$^{1}$Living Matter Department, AMOLF, Science Park 104, 1098 XG
Amsterdam}

\affiliation{$^{2}$Departments of Chemical \& Biomolecular Engineering, Chemistry,
and Physics \& Astronomy, Rice University, Houston, TX 77005, USA}

\affiliation{$^{3}$Center for Theoretical Biophysics, Rice University, Houston,
TX 77030, USA }

\affiliation{$^{4}$Department of Physics and Astronomy, Vrije Universiteit, 1081
HV Amsterdam, The Netherlands}

\maketitle
Biopolymers form the scaffolds of life, providing rigidity to both
cells and the extracellular matrix \citep{Bausch2006,Broedersz2014c,Meng2017}.
An important characteristic of intra- and extracellular biopolymers
\citep{Gittes1993,Lin2010,Piechocka2016} is their high bending rigidity
relative to most synthetic polymers. This feature leads to a competition
between entropic and energetic effects that results in a range of
material properties not captured by traditional polymer physics. One
such property is the highly nonlinear elastic response of biopolymer
networks, in which the shear rigidity can increase by orders of magnitude
upon strains of by only a few percent \citep{Gardel2004}. For permanently
crosslinked semiflexible polymer networks, this phenomenon is well
accounted for by the compliance due to transverse bending fluctuations
that become suppressed under a load \citep{F.C.MacKintoshJ.Kas1995,Statics2005}. 

The interactions of biopolymers are also more complex than for traditional
polymer materials. An example is the transient cross-linking by specialized
crosslinker proteins that takes place in the actin cytoskeleton of
the cell, which causes stress relaxation on timescales much longer
than the typical cross-linker unbinding time \citep{Mayer2010,Fischer-Friedrich2016}.
The resulting viscoelastic flow does not follow a simple Maxwell model
with a single relaxation time, but instead follows power law behavior
characteristic of a broad spectrum of relaxation times \citep{Lieleg2008,Broedersz2010}. 

Here, we show that the nonlinear response of transiently crosslinked
actin networks exhibits an unexpectedly slow stress relaxation, resembling
the dynamics of soft glassy systems \citep{Sollich1997}. 
As reported in Ref.\ \citep{Fabry2001} 
and a large body of follow-up work on cell rheology \citep{Bursac2005,Desprat2005,Deng2006,Trepat2007,Trepat2008,Kollmannsberger2011}, 
the shear modulus of cells is characterized by a shear modulus that increases as a weak power-law of frequency, 
with exponents as low as 0.1, for which the term soft glassy rheology has been used.
Interestingly, in contrast with prior models of soft glassy rheology of cytoskeletal networks
\citep{Semmrich2007,Kroy2007,Meng2018}, here we show that exponents below 0.5 are only observed
in the nonlinear regime. We show that the time- and stress-dependent
response of actin networks is consistent with a model that accounts
for both the nonlinear stiffening \citep{F.C.MacKintoshJ.Kas1995,Gardel2004}
and transient crosslinking \citep{Broedersz2010} of semi-flexible
polymers. Our results can provide an explanation for the many prior reports
of slow relaxation and near solid-like viscoelastic response in reconstituted
cytoskeletal networks \citep{Semmrich2007,Lieleg2011} and in living
cells \citep{Fabry2001,Bursac2005,Desprat2005,Deng2006,Trepat2007,Trepat2008,Kollmannsberger2011}.
While these phenomena have been discussed in the context of phenomenological
soft glassy rheology \citep{Fabry2001,Bursac2005,Semmrich2007,Kroy2007,Trepat2007,Lieleg2011},
a more microscopic mechanism has been lacking. The present work suggests
that the glassy dynamics in the cytoskeleton are a natural consequence
of transient cross-links, combined with prestress. 

\begin{figure}
\begin{minipage}[b]{0.5\columnwidth}%
\subfloat[\label{fig:Schematic_dynamically_crosslinked}]{\includegraphics[width=1\columnwidth]{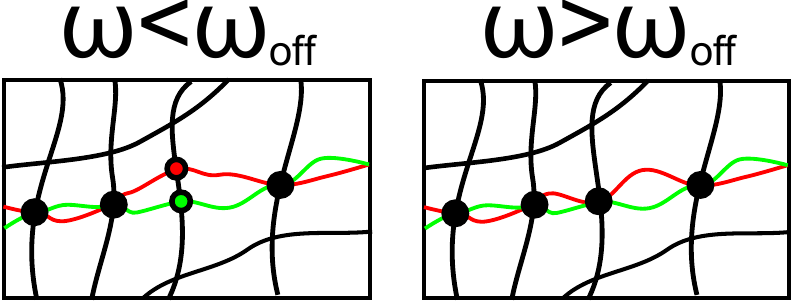}

}

\subfloat[\label{fig:Phase-diagram_dynamically_crosslinked}]{\includegraphics[width=1\columnwidth]{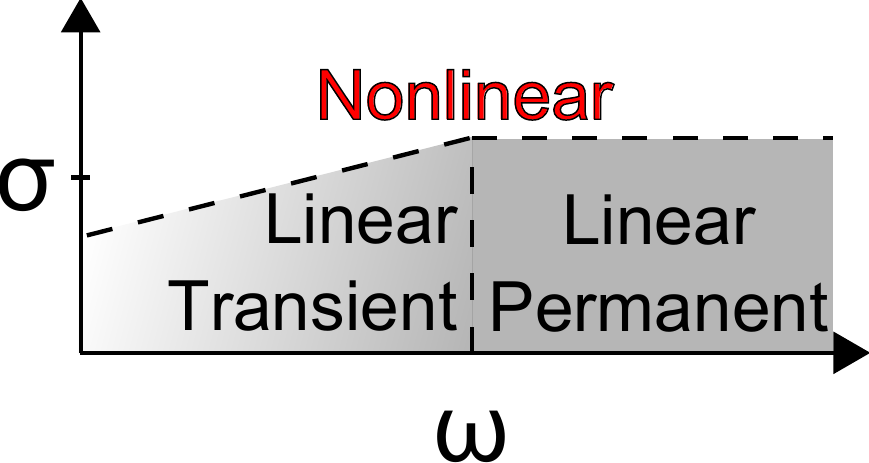}

}%
\end{minipage}\subfloat[\label{fig:Linear-rheology-of}]{\includegraphics[width=0.5\columnwidth]{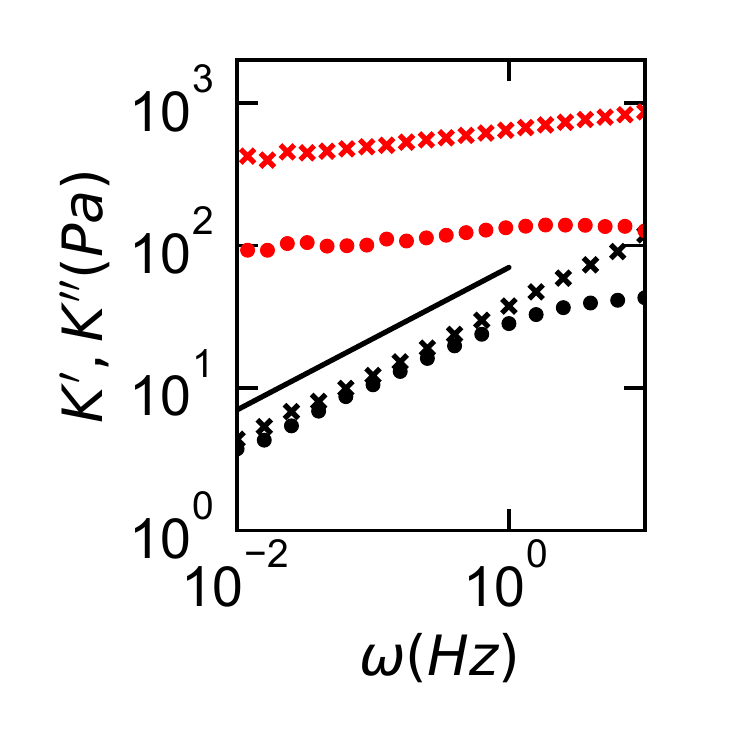}

}

\caption{\textbf{Time-dependent viscoelastic response of transiently crosslinked
semiflexible polymer networks.} a) Schematic showing actin filaments
(black lines) connected by crosslinkers (black dots). Left: filament
before and after a crosslinker remodeling event (green and red). Right:
such events are unlikely for frequencies larger than the crosslinker
unbinding rate $\omega_{\textrm{off}}$. b) Schematic regime diagram
showing three different viscoelastic regimes as a function of frequency
and applied stress. At low stress, in the linear regime, networks
either behave as permanent networks exhibiting a plateau or crosslink
kinetics lead to a frequency-dependent transient regime. Our work
shows that beyond an onset stress, a single length scale that is nonlinear
in both stress and frequency governs the mechanics. This onset stress
decreases for frequencies below the crosslinker unbinding rate as
the effective crosslinker distance decreases. c) The storage (crosses)
and loss moduli (circles) of a crosslinked actin network against frequency
in the absence of prestress (black) and for $8$ Pa prestress (red).
The line indicates a 1/2 power law.\label{fig:Phase-diagram-and-viscous}}
\end{figure}

Using small amplitude oscillatory rheology, we measure the storage
(crosses) and loss moduli (circles) in the absence of prestress (black
data points) as a function of frequency for reconstituted actin networks,
crosslinked by the dynamic linker $\alpha$-actinin-4 (ACTN4), a prominent
crosslinker in human cells \citep{Thomas2017,Feng2018} {[}Fig. \ref{fig:Linear-rheology-of}{]}.
Qualitatively consistent with previous experiments and modeling \citep{Broedersz2010},
we find a power law frequency dependence of the moduli at frequencies
below $1$ Hz (black line) with an exponent close to 1/2. The $\nicefrac{1}{2}$
exponent reflects the broad spectrum of relaxation times from the
unbinding and rebinding of multiple crosslinkers along a filament
{[}Fig. \ref{fig:Schematic_dynamically_crosslinked}{]} \citep{Broedersz2010}.
The viscous modulus becomes less frequency dependent for higher frequencies
($>1$ Hz), and is expected to peak at the crosslinker unbinding
rate (not observed here and therefore likely beyond $10\,Hz)$ as
crosslinker unbinding becomes increasingly unlikely \citep{Broedersz2010}. 

Next, to probe the nonlinear response, we measure the differential
modulus $\partial\sigma/\partial\gamma$, where $\sigma$ is the shear
stress and $\gamma$ is strain. We do so by superimposing a small
amplitude oscillation on $8$ Pa prestress (red data points in Fig.
\ref{fig:Linear-rheology-of}). We find that both real (storage) and
imaginary (loss) moduli are larger in the presence of prestress by
1-2 orders of magnitude than in the absence of prestress over the
entire frequency range. We attribute this change to the stress stiffening
response of semiflexible polymer networks to suppression of filament
bending fluctuations \citep{F.C.MacKintoshJ.Kas1995,Statics2005}.
More surprisingly, we find that both the storage and loss moduli are
significantly less frequency-dependent in the presence of prestress
than the $\nicefrac{1}{2}$ power law observed in the absence of prestress.

\begin{figure}
\subfloat[\label{fig:diff_el_vs_stress_freq}]{\includegraphics[width=0.5\columnwidth]{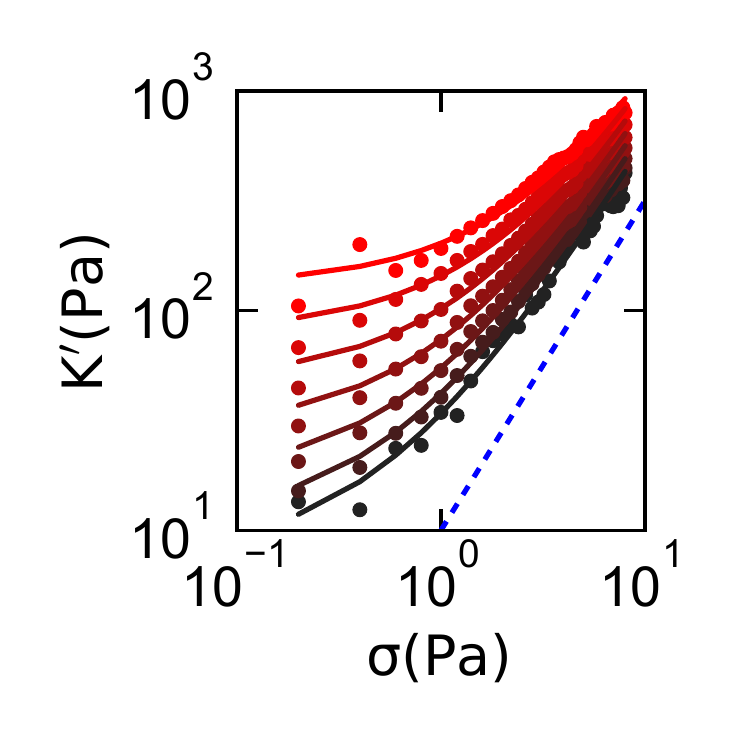}}\subfloat[\label{fig:diff_vis_vs_stress}]{\includegraphics[width=0.5\columnwidth]{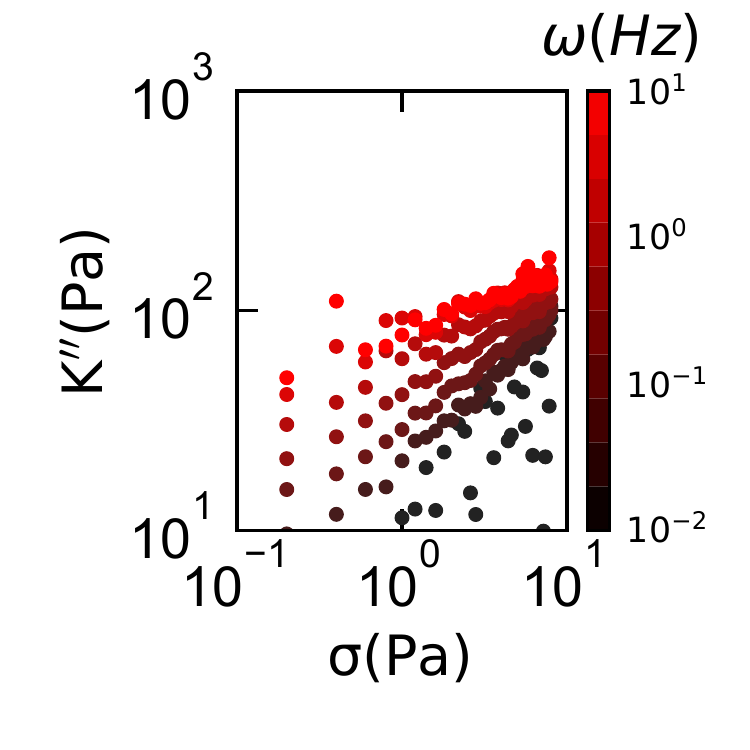}}

\caption{\textbf{Time-dependent stiffening.} The differential storage (a) and
loss (b) moduli of an ACTN4 crosslinked actin network are plotted
against the applied stress and color coded as a function of frequency
from $0.01$ Hz in black to $5$ Hz in red in 7 logarithmically
spaced steps. The solid lines represent fits to Eq. (\ref{eq:stiffening}).
The blue dashed line represents the $\nicefrac{3}{2}$ power law characteristic
of \emph{permanently} crosslinked networks \citep{Gardel2004}.\label{fig:Frequency-dependent-stiffening}}
\end{figure}

\begin{figure}
\subfloat[\label{fig:onset_stiffening_vs_freq}]{\includegraphics[width=0.5\columnwidth]{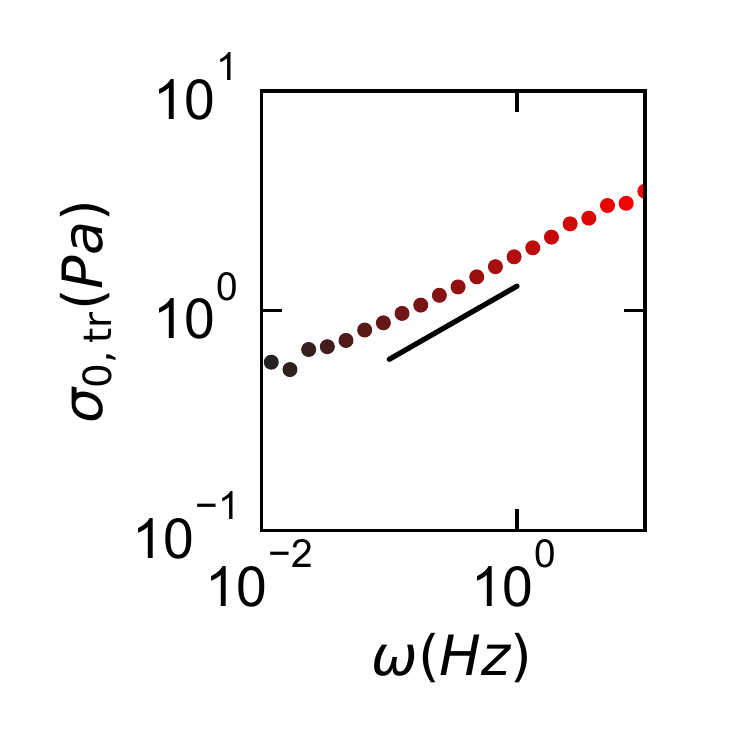}}\subfloat[\label{fig:collapse_curve}]{\includegraphics[width=0.5\columnwidth]{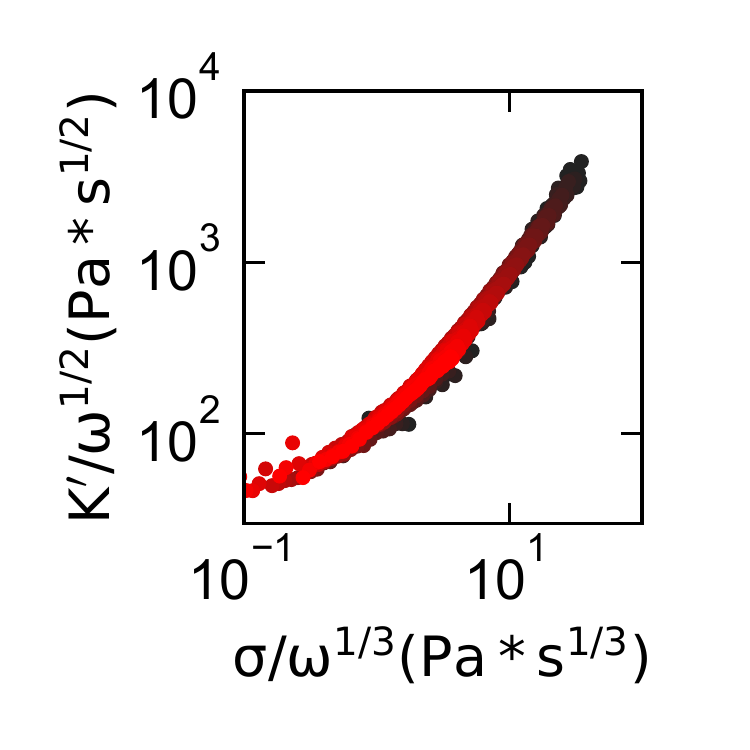}}\caption{\textbf{Mastercurve behavior of the time and stress dependent viscoelastic
behavior of actin networks.} a) The onset stress for stiffening of
actin networks follows a $\nicefrac{1}{3}$ power law dependence on
frequency (black line), consistent with Eq. (\ref{eq:onset_stress}).
b) Stress-stiffening curves over all frequencies can be collapsed
onto a single mastercurve using Eq. (\ref{eq:K_transient}). The color
coding is identical to Fig. \ref{fig:Frequency-dependent-stiffening}.\label{fig:collapse_and_onset_stress}}
\end{figure}

To find out the origin of the stress-dependent changes in the time-dependent
rheology, we systematically vary the prestress over a range from $0.1$
to $8.0$ Pa with a superimposed small amplitude oscillation at different
frequencies ($\omega=0.01...10$ Hz). We find that both the differential
storage and loss moduli increase as a function of prestress over all
frequencies {[}Fig. \ref{fig:Frequency-dependent-stiffening}{]}.
This increase is consistent with an asymptotic $\sigma^{\nicefrac{3}{2}}$
power-law stress stiffening (indicated by the blue dashed line), as
previously identified both experimentally and theoretically for semiflexible
polymer networks at high $\sigma$ \citep{Gardel2004,Lin2010}. To
test the agreement with the model more quantitatively, we fit the
differential storage modulus at each frequency to the following cross-over
function: 

\begin{equation}
K'=G\cdot(1+\nicefrac{\sigma}{\sigma_{0,\textrm{tr}}})^{\nicefrac{3}{2}},\label{eq:stiffening}
\end{equation}
where $G(\omega)$ is the linear storage modulus and $\sigma_{0,\textrm{tr}}$
is the characteristic stress for the onset of stiffening at a given
frequency. Remarkably, although the model for the $\sigma^{3/2}$
stiffening was developed and confirmed previously for the nonlinear
elasticity of \emph{permanently} crosslinked networks, it accurately
captures the nonlinear elastic response of transiently crosslinked
actin networks as well. However, whereas the onset stress for nonlinearity
of permanently crosslinked networks is independent of frequency \citep{Gardel2004,Lin2010},
we find that this onset systematically increases with frequency for
transiently crosslinked semiflexible polymer networks {[}Fig. \ref{fig:onset_stiffening_vs_freq}{]}. 

In order to capture both the frequency and the stress dependence of
the shear moduli, we propose a model in which we combine the transient
nature of the crosslinkers with the nonlinear force-extension behavior
of the semiflexible actin filaments. For permanently crosslinked networks,
the storage modulus in the linear regime is dependent on the distance
between crosslinkers, $l_{o}$ \citep{F.C.MacKintoshJ.Kas1995}:
\begin{equation}
G_{0}\sim\rho\kappa l_{p}/l_{0}^{3},
\end{equation}
where $l_{p}$ is the persistence length of the filament, $\rho$
the filament length density per volume and $\kappa$ the bending rigidity
of the filament. In the nonlinear regime, the modulus becomes independent
of the distance between crosslinkers, but is defined by the length
scale beyond which bending wavelengths are suppressed due to the filament
axial tension $\tau$ \citep{Broedersz2014c}:
\begin{equation}
l_{\tau}\sim\sqrt{\nicefrac{\kappa}{\tau}}.
\end{equation}
As filament axial tension increases with the applied stress, $\sigma\sim\rho\tau$,
the relevant bending wavelengths become smaller and the storage modulus
increases nonlinearly with the applied stress:
\begin{equation}
K\sim\rho\kappa l_{p}/l_{\tau}^{3}\sim\frac{\rho\kappa l_{p}}{l_{0}^{3}}(\nicefrac{\sigma}{\sigma_{0}})^{\nicefrac{3}{2}},
\end{equation}
where $\sigma_{0}\sim\rho\kappa/l_{0}^{2}$ is the threshold stress
at which the network begins stiffening as the typical filament bending
wavelength decreases below the typical crosslinker distance. 

The important difference between transient networks and permanently crosslinked ones is that
the effective crosslink distance increases with time as longer wavelength bending modes relax due to crosslink 
unbinding and rebinding \citep{Broedersz2010}. This can be captured by an effective crosslink distance
\begin{equation}
l_{\textrm{tr}}\sim\omega^{-\nicefrac{1}{6}}>l_{0}\mbox{ for }\omega<\omega_{\textrm{off}},\label{eq:distance_frequency}
\end{equation}
which leads to a $\omega^{0.5}$ dependence of the shear modulus, 
as reported experimentally and theoretically in Ref.\ \citep{Broedersz2010}.
Here, $\omega_{\textrm{off}}$ is the crosslink unbinding rate.
As a result, the onset for stress stiffening now depends on frequency
according to 
\begin{equation}
\sigma_{0,\textrm{tr}}\sim\rho\kappa/l_{\textrm{tr}}^{2}\sim\omega^{\nicefrac{1}{3}},\label{eq:onset_stress}
\end{equation}
consistent with Fig. \ref{fig:onset_stiffening_vs_freq}. In order
to capture both transient and non-transient regimes, we let
\begin{equation}
l_{\mathrm{tr}}=l_{0}\left(1+\sqrt{\nicefrac{\omega_{\textrm{off}}}{\omega}}\right)^{\nicefrac{1}{3}}.\label{eq:distance_time}
\end{equation}
Strictly speaking, this is correct in the asymptotic plateau ($\omega\gg\omega_{\textrm{off}}$)
and transient ($\omega\ll\omega_{\textrm{off}}$) regime, while it
is only approximate in the crossover regime at intermediate frequencies.

Similarly, in order to approximate the crossover from the linear to
the nonlinear regime, we let
\begin{equation}
l_{\tau}=l_{\mathrm{tr}}(1+(\sigma/\sigma_{\textrm{0,tr}}))^{-1/2}.\label{eq:distance_stress}
\end{equation}
Again, strictly speaking, this is correct for linear ($\sigma\ll\sigma_{\textrm{0,tr}}$)
and highly nonlinear ($\sigma\gg\sigma_{\textrm{0,tr}}$) regimes,
although we show below that it approximates well the behavior of actin
networks over the entire experimentally accessible range of stress. 

\begin{figure}
\subfloat[\label{fig:fit_el_vs_freq}]{\includegraphics[width=0.5\columnwidth]{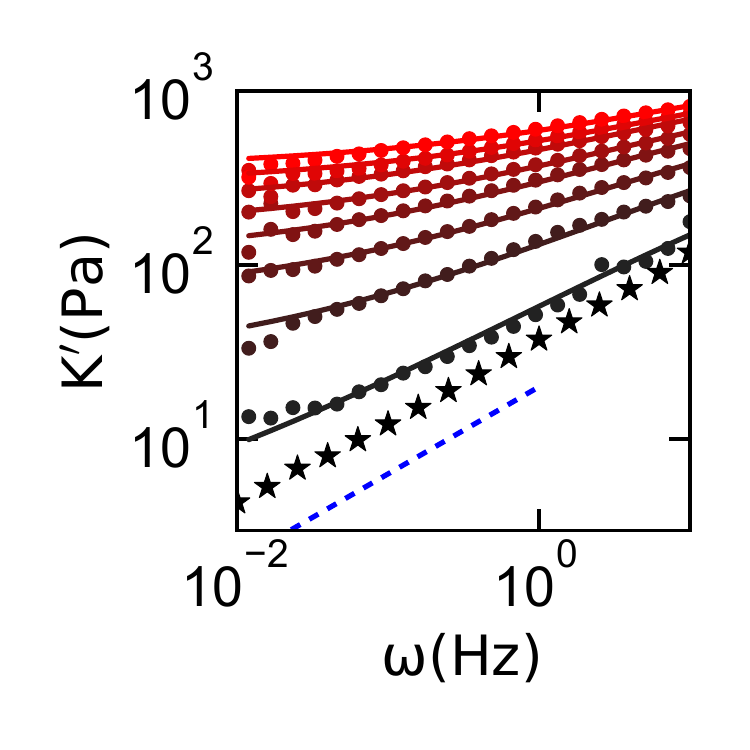}}\subfloat[\label{fig:diff_vis_vs_freq}]{\includegraphics[width=0.5\columnwidth]{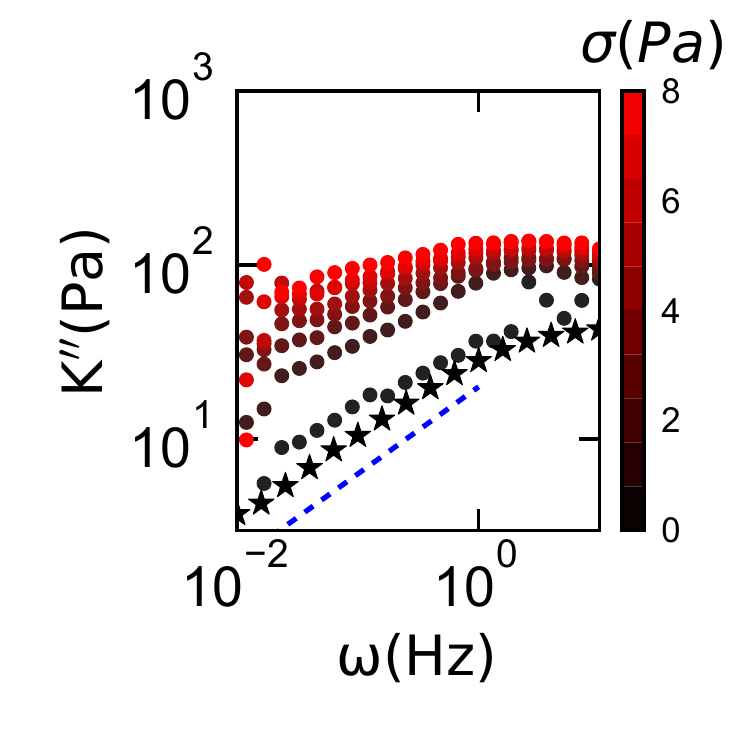}}

\subfloat[\label{fig:power_law}]{\includegraphics[width=0.5\columnwidth]{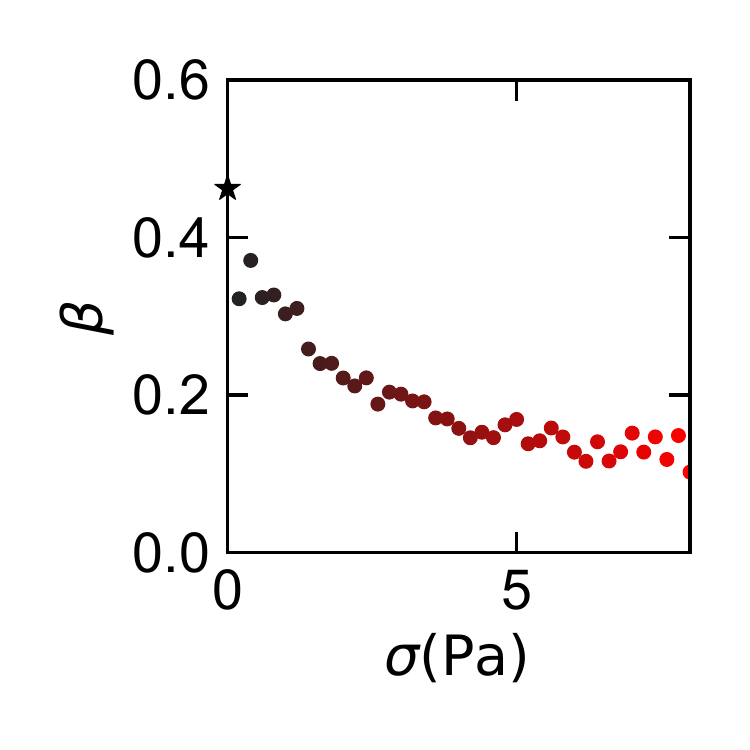}}\subfloat[\label{fig:power_law_pref}]{\includegraphics[width=0.5\columnwidth]{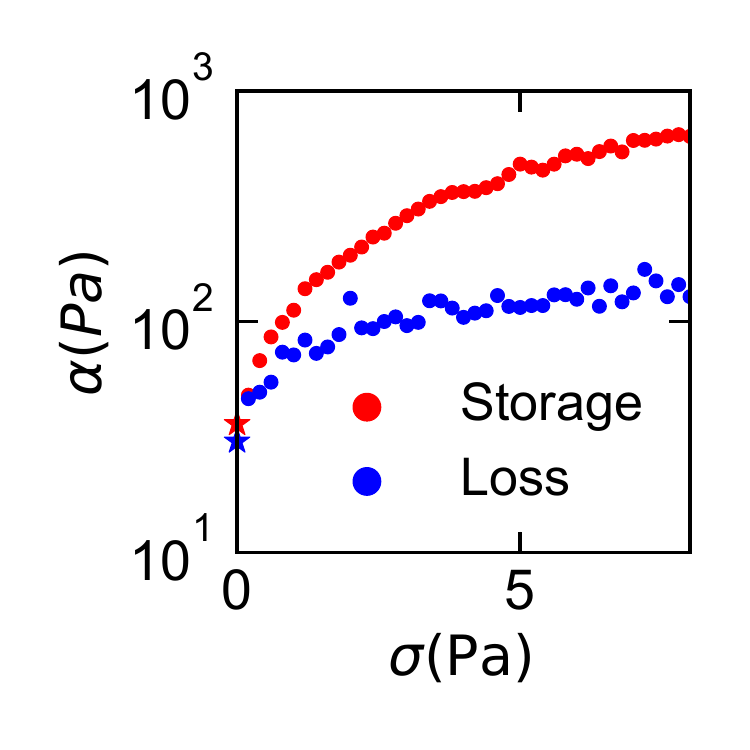}}

\caption{\textbf{Frequency dependence of stressed actin networks.} The differential
storage (a) and loss (b) moduli of an ACTN4 crosslinked actin network
are plotted against the applied frequency and color coded as a function
of prestress from $0.2$ Pa in black to $7.2$ Pa in red with $1$ Pa
steps. The stars are data in the absence of prestress. The blue line
represents the $\nicefrac{1}{2}$ power law characteristic of networks
\emph{in the absence of stress.} \citep{Broedersz2010}. The solid
lines represent fits to $K'(\sigma,\omega)$ according to Eq. (\ref{eq:K_transient})
(see Fig. S1 for the fitting parameters \citep{SI}). The differential moduli
are fitted between 0.01 and 1 Hz with an empirical power law $\alpha(\sigma)\cdot\omega^{\beta(\sigma)}$.
The exponent $\beta$ (c) and prefactor $\alpha$ (d) are shown as
a function of stress.\label{fig:Glassy-dynamics-in}}
\end{figure}

The resulting expression for $K'$ is
\begin{equation}
K'\sim\rho\frac{\kappa l_{p}}{l_{0}^{3}}\frac{[1+(\nicefrac{\sigma}{\sigma_{0,\textrm{tr}}})]^{3/2}}{(1+\sqrt{\nicefrac{\omega_{\textrm{\textrm{off}}}}{\omega}})}.\label{eq:K_transient}
\end{equation}
This model accurately describes the observed trends in the nonlinear
rheology of actin networks. Firstly, the theory predicts a $\omega^{\nicefrac{1}{3}}$
scaling of onset stress for nonlinearity (Eq. (\ref{eq:onset_stress})
), consistent with our experimental data {[}Fig. \ref{fig:onset_stiffening_vs_freq}{]}.
Secondly, using Eq. (\ref{eq:K_transient}), we successfully collapse
all stress-stiffening data {[}Fig. \ref{fig:collapse_curve}{]}. Lastly,
we use Eq. (\ref{eq:K_transient}) to accurately fit the differential
storage modulus as a function of frequency over all prestresses {[}Fig.
\ref{fig:fit_el_vs_freq}{]}. Interestingly, using $\omega_{\textrm{off}}$
as a free parameter, we find that the characteristic frequency decreases
as the applied stress is raised {[}Fig. S1{]}. This result is consistent
with earlier rheological measurements on networks crosslinked by ACTN4
\citep{Yao2013}, and suggests catch bonding \citep{Chakrabarti2016}
where the crosslinker unbinding rate decreases with force \citep{Schiffhauer2016}. 

We find that the elastic response of transiently crosslinked actin
networks is well captured by a simple model that combines prior models
for the linear viscoelasticity of transient gels and the nonlinear
elasticity of permanent networks. Key to our model is a single length
scale defined by Eqs. (\ref{eq:distance_time},\ref{eq:distance_stress})
that characterizes the upper limit of fluctuating wavelengths. This
length depends on both time/frequency and stress. Together, these
effects result in a frequency dependence that becomes weaker with
increasing stress {[}Fig. \ref{fig:Glassy-dynamics-in}a,b{]}. The
$K\sim\omega^{\nicefrac{1}{2}}$ power law predicted by Ref. \citep{Broedersz2010}
is only observed in the absence of prestress {[}Fig. \ref{fig:Linear-rheology-of}{]}.
We quantify the dependence of stress-stiffening on stress by fitting
the data with an empirical power law $K(\sigma,\omega)=\alpha(\sigma)\cdot\omega^{\beta(\sigma)}$
commonly used in the cell rheology literature \citep{Fabry2001,Desprat2005,Deng2006,Trepat2007,Kollmannsberger2011}.
We find that the prefactor $\alpha$ increases for the loss modulus
and, even more steeply, for the storage modulus as the network stiffens,
such that the network becomes more solid-like with increasing stress
{[}Fig. \ref{fig:power_law_pref}{]}. We also find that the exponent
$\beta$ decreases from 0.5 in the absence of stress to 0.1 at 8 Pa
{[}Fig. \ref{fig:power_law}{]}. This is in contrast to the mechanics
of permanently crosslinked networks \citep{F.C.MacKintoshJ.Kas1995,Statics2005}
that are time-independent except at very high frequencies, typically
beyond 100 Hz, where the viscous drag of the fluid controls filament
relaxation \citep{Gittes1997,Morse1998,Gittes1998,Koenderink2006}.
In that regime, an exponent $\beta$ of $\nicefrac{3}{4}$ is expected,
but this can decrease to $\nicefrac{1}{2}$ under stress \citep{Mizuno2007}.
Recently, Ref. \citep{Meng2018} proposed a model for the nonlinear
response of transient semiflexible networks, but no specific relationship
between the stress and the exponent governing the time dependence
was predicted.

Other work on stressed dynamically crosslinked actin networks has
focused on the effect of force-induced linker (un)binding \citep{Lieleg2009,Heussinger2012,Yao2013},
sliding of crosslinkers along filaments \citep{Kroy2012b,Plagge2016}
and the effect of shear-induced filament alignment \citep{Majumdar2018}.
While our minimal model does not include such effects, it is able
to accurately capture the stress and frequency dependence of the nonlinear
elastic response of actin networks. In future work, it would be interesting
to include the additional microscopic effects mentioned above \citep{Majmudar2005,Lieleg2009,Heussinger2012,Kroy2012b,Yao2013,Plagge2016},
as well as to quantitatively understand the differential loss modulus,
$K"$, for example by using detailed network simulations of transiently
connected semiflexible polymers \citep{Muller2014b}.

We find that stressed semiflexible polymer networks exhibit power
law dynamics with a small exponent ($\beta\sim0.1$). Remarkably,
mechanical experiments on living cells have revealed similar power
law dynamics \citep{Fabry2001,Desprat2005,Bursac2005,Deng2006,Trepat2007,Trepat2008,Kollmannsberger2011}.
These mechanical properties are reminiscent of observations on a range
of disordered systems close to the glass transition \citep{Sollich1997}.
The soft glassy rheological properties found in cells have been phenomenologically
described by assuming a broad distribution of microscopic relaxation
timescales \citep{Fabry2001,Semmrich2007,Kroy2007,Lieleg2011}. Whilst
this phenomenological description can account for the experimental
data \citep{Fabry2001,Desprat2005,Deng2006,Semmrich2007,Trepat2007,Trepat2008,Kollmannsberger2011,Lieleg2011},
it offers no insight into the microscopic processes governing these
dynamics. Here we suggest that the glassy dynamics are a natural consequence
of transient cross-links, combined with prestress. This mechanism is different from the microscopic mechanism underlying soft, glassy rheology in systems such as colloidal gels, where particle density, rather than prestress, controls the stress relaxation exponent \citep{berthier2011theoretical}.

Myosin motor-driven contractility is a likely source for such prestress
in the cell \citep{Mizuno2007,Koenderink2009d}. Consistent with our
results, experiments on cells have revealed that the power law exponent
of the frequency dependent shear moduli decreases with the internal
stress generated by actomyosin contractility \citep{Kollmannsberger2011}.
Remarkably though, whereas the reconstituted networks become more
solid-like with prestress {[}Fig. \ref{fig:power_law_pref}{]}, mechanical
experiments on cells have shown that the loss modulus increases more
rapidly than the storage modulus as a function of myosin-driven tension
\citep{Kollmannsberger2011}. We speculate that external stress as
imposed in our rheological experiments differs from internal stresses
generated by myosin motors, as motors not only cause contractility
but also fluidize networks via filament sliding \citep{Liverpool2001,Humphrey2002d}
and severing \citep{Murrell2012,McCall2017}.

\section*{{\normalsize{}Acknowledgements}}

We thank William Brieher and Vivian Tang for the kind gift of purified
ACTN4, Marjolein Kuit-Vinkenoog for actin purification, Bela Mulder
for critical reading of the manuscript and Chase Broedersz for useful
discussions. This work is part of the research program of the Netherlands
Organisation for Scientific Research (NWO). We gratefully acknowledge
financial support from an ERC Starting Grant (335672-MINICELL). F.C.M.
was supported in part by the National Science Foundation (Grants PHY-1427654
and DMR-1826623).

\bibliographystyle{unsrt}

\end{document}